\documentclass[aps,floats,floatfix,superscriptaddress,twocolumn]{revtex4}

\usepackage{amsfonts,amsmath,amssymb,hyperref}
\usepackage[pdftex]{graphicx}
\usepackage{color}
\usepackage{times}

\renewcommand{\leq}{\leqslant}
\renewcommand{\geq}{\geqslant}

\begin{document}

\title{Emergence of Soft Communities  from Geometric Preferential Attachment}

\author{Konstantin Zuev}
\affiliation{Department of Physics, Northeastern University, Boston, MA 02115, USA}

\author{Mari{\'a}n Bogu{\~n}{\'a}}
\affiliation{Departament de F{\'\i}sica Fonamental, Universitat de Barcelona, Mart\'i i Franqu\`es 1, 08028 Barcelona, Spain}

\author{Ginestra Bianconi}
\affiliation{School of Mathematics, Queen Mary University of London, London E1, 4SN, UK}

\author{Dmitri Krioukov}
\affiliation{Department of Mathematics and Department of Electrical \& Computer Engineering, Northeastern University, Boston, MA 02115, USA}

\begin{abstract}
All real networks are different, but many have some structural properties in common. There seems to be no consensus on what the most common properties are, but scale-free degree distributions, strong clustering, and community structure are frequently mentioned without question. Surprisingly, there exists no simple generative mechanism explaining all the three properties at once in growing networks. Here we show how latent network geometry coupled with preferential attachment of nodes to this geometry fills this gap. We call this mechanism {\em geometric preferential attachment} (GPA), and validate it against the Internet. GPA gives rise to {\em soft communities} that provide a different perspective on the community structure in networks. The connections between GPA and cosmological models, including inflation, are also discussed.
\end{abstract}

\maketitle

\section{Introduction}
One of the fundamental problems in the study of complex networks~\cite{Dorogovtsev_book1,NBW,BBB,Dorogovtsev_book2,NetsCrowdsMarkets} is to identify
evolution mechanisms that shape the structure and dynamics of large
real networks such as the Internet, the world wide web, and various biological and
social networks. In particular, how do complex networks grow so that
many of them are scale-free and have strong clustering and non-trivial community structure?
The preferential attachment (PA) mechanism~\cite{PA,Krapivsky,Dorogovtsev2}, where new \textit{connections} are made
preferentially to more popular nodes, is widely accepted as the
plausible explanation for the emergence of the scale-free structures
(i.e.\ the power-law degree distributions) in large networks.
PA has been empirically validated for many real growing
networks~\cite{Newman,BarabasiJeongNeda,Vazquez,Jeong} using statistical analysis of a sequence of network
snapshots, demonstrating that it is indeed a key element of network evolution. Moreover, there is some evidence that the evolution of the community graph --- a graph where nodes represent communities and links refer to members shared by two communities --- is also driven by PA~\cite{Pollner}.

Nevertheless, PA alone cannot explain two other empirically observed universal properties of complex networks: strong clustering~\cite{WS} and significant community structure~\cite{GirvanNewman}. Namely, in synthetic networks generated by standard PA, clustering
is asymptotically zero~\cite{Bollobas} and there are no communities~\cite{Fortunato}. To resolve the zero-clustering problem, several modifications of the original PA mechanism  have been proposed~\cite{Dorogovtsev3,Klemm,Vazquez2,Jackson}. To the best of our knowledge, however, none of these models capture all three fundamental properties of complex networks: heavy-tail degree distribution, high clustering, and community structure.

In social networks, the presence of communities, that might represent node clusters based on certain social factors such as economic status or political beliefs, is intuitively expected. A remarkable observation~\cite{GirvanNewman,www,metabolic,biochemical,financial,metabolic2} is that many other networks, including food webs, the world wide web, metabolic, biochemical, and financial networks, also admit a reasonable division into informative communities. Since that discovery, community detection has become one of the main tools for the analysis and understanding of network data~\cite{Fortunato,Danon}. 

Despite an enormous amount of attention to community detection algorithms and their efficiency, there were very few attempts to answer a more fundamental question: what is the actual mechanism that induces community structure in real networks? For social networks, where  there is a strong relationship between a high concentration of triangles and the existence of community structure~\cite{NP}, triadic closure~\cite{Rapoport} has been proposed as a plausible mechanism for generating communities~\cite{Bianconi}. It was also shown by means of a simple agent-based acquaintance model that a large-scale community structure can emerge from the underlying social dynamics~\cite{Bhat}. There also exist other contributions in this direction, where proposed mechanisms and generative models are specifically tailored for social networks~\cite{Marian,Toivonen,Lambiotte,Lambiotte2}.

Here we show how latent network geometry coupled with preferential attachment of \textit{nodes} to this geometry induces community structure as well as power-law degree distributions and strong clustering. We prove that these universal properties of complex networks naturally emerge from the new mechanism that we call \textit{geometric preferential attachment} (GPA), without appealing to the specific nature (e.g.\ social) of networks. Using the Internet as an example, we demonstrate that GPA generates networks that are in many ways similar to real networks.

\section{Results}
\subsection{Geometric Preferential Attachment}

In growing networks the concept of popularity that PA exploits is just one aspect of node attractiveness; another important aspect is similarity~\cite{PSO}.  Namely, if nodes are similar (``birds of feather''), then they have a higher chance of being
connected (``flock together''), even if they are not popular. This effect, known as homophily in social sciences~\cite{McPherson}, has been observed in many  real networks of various nature~\cite{Redner,Watts}.

The GPA mechanism utilizes the idea that both popularity and similarity are important. We take the node birth time $t=1,2,\ldots$ as a proxy for node's popularity: all other things being equal, the older the node (i.e.\ the smaller $t$), the more popular it is. The similarity attribute of node $t$ is modeled by a random variable $\theta_t$ distributed over a circle $\mathbb{S}^1$ that abstracts the ``similarity'' space. One can think of the similarity space as an image of a certain projection $p: \mathcal{A}\rightarrow\mathbb{S}^1$ from a space of unknown or not easily measurable  attributes $(a^1,\ldots,a^k)\in\mathcal{A}$ of nodes. For social networks, these attributes could be political beliefs, education, and social status, whereas for biological networks, $\{a^i\}$ may represent chemical properties of metabolites or geometric properties of protein shapes. While the absolute value of the similarity coordinate $\theta_t=p(a_t^1,\ldots,a_t^k)$ does not have any specific meaning, the angular distance $\theta_{st}=\pi-|\pi-|\theta_s-\theta_t||$ quantifies the similarity between two nodes. Upon its birth, a new node $t$ connects to an existing node $s<t$ if $s$ is both popular enough and similar to $t$, that is if $s^\beta\theta_{st}$ is small, where $\beta\in[0,1]$ is a parameter that controls the relative contributions of popularity and similarity.

The described rule for establishing new connections admits a simple geometric interpretation which is very useful for analytical treatment of the model. Let us define the radial coordinate of node $s$ at time $s$ as $r_s=2\ln s$, and let it grow
with time, so that at time $t>s$ it is $r_s(t)=\beta r_s + (1-\beta)r_t$. The distance $x_{st}$ between two points in the hyperbolic
plane of curvature $K=-1$ with polar coordinates
$(r_s(t),\theta_s)$ and $(r_t,\theta_t)$ is approximately~\cite{Bonahon} $x_{st}=r_s(t)+r_t+2\ln\frac{\theta_{st}}{2}=2\ln\left(\frac{s^\beta	t^{2-\beta}\theta_{st}}{2}\right)$. Since for any given $t$, the sets of nodes $s<t$ that minimize $s^\beta\theta_{st}$ and $x_{st}$ are the same, new nodes simply connect to the hyperbolically closest existing nodes. Note that the increase of the radial coordinate $r_s(t)$
decreases the effective age of the node, and thus models the effect of
popularity fading observed in many real networks~\cite{Raan}.

But how do new nodes find their positions in this similarity space? The main assumption of our model is that the hidden attribute space $\mathcal{A}$ of a growing network is likely to contain ``hot'' regions (e.g.\ of
human activity), and that the hotter the region, the more attractive it is for new nodes.
Hot regions can for instance represent some hot areas in science.  When these regions are projected onto the similarity space $\mathbb{S}^1$, the hotness
manifests itself by a higher node density, more scientists working in a hot area. The higher attractiveness of a hot region is then modeled by placing a new node in this region with the higher probability, the hotter this region is, i.e.\ the higher the node density in it. That is, new scientists are expected to begin their careers working in hot areas where many existing scientists are already active, versus jumping onto some obscure developments that nobody understands. Therefore the higher the node density in a particular section of our similarity space $\mathbb{S}^1$, the higher the probability that a new node is placed in this section. Intuitively we would expect that this process should lead to heterogeneous distributions of node coordinates in the similarity space. This intuition is confirmed by empirical results: if we map real networks to their hyperbolic spaces~\cite{SustainingInt,HyperMap}, we observe that the resulting empirical angular node density is not uniform (e.g.\ see Fig.~5(a)), and nodes tend to cluster into tight communities.  In the Internet, for example, these communities are groups of Autonomous Systems belonging to the same country.

There are many ways to implement this general idea. For a variety of reasons we found that the most natural and consistent one is as follows. First we define the attractiveness of any location $\varphi\in\mathbb{S}^1$ for a new node $t$ with radial coordinate $r_t$ as the number of existing nodes $s<t$ lying in the hyperbolic disk
$D_\varphi(r_t)$ of radius $r_t$ centered at
($r_t$,$\varphi$). The higher the attractiveness of a location $\varphi$, the higher the probability that a new node $t$ will chose this location as its place $\theta_t=\varphi$ in the similarity space. We refer to this mechanism as the geometric preferential attachment (GPA) of nodes to the similarity space. This mechanism is illustrated in Fig.~1.

\begin{figure}[t]
	\centerline{\includegraphics[width=90mm]{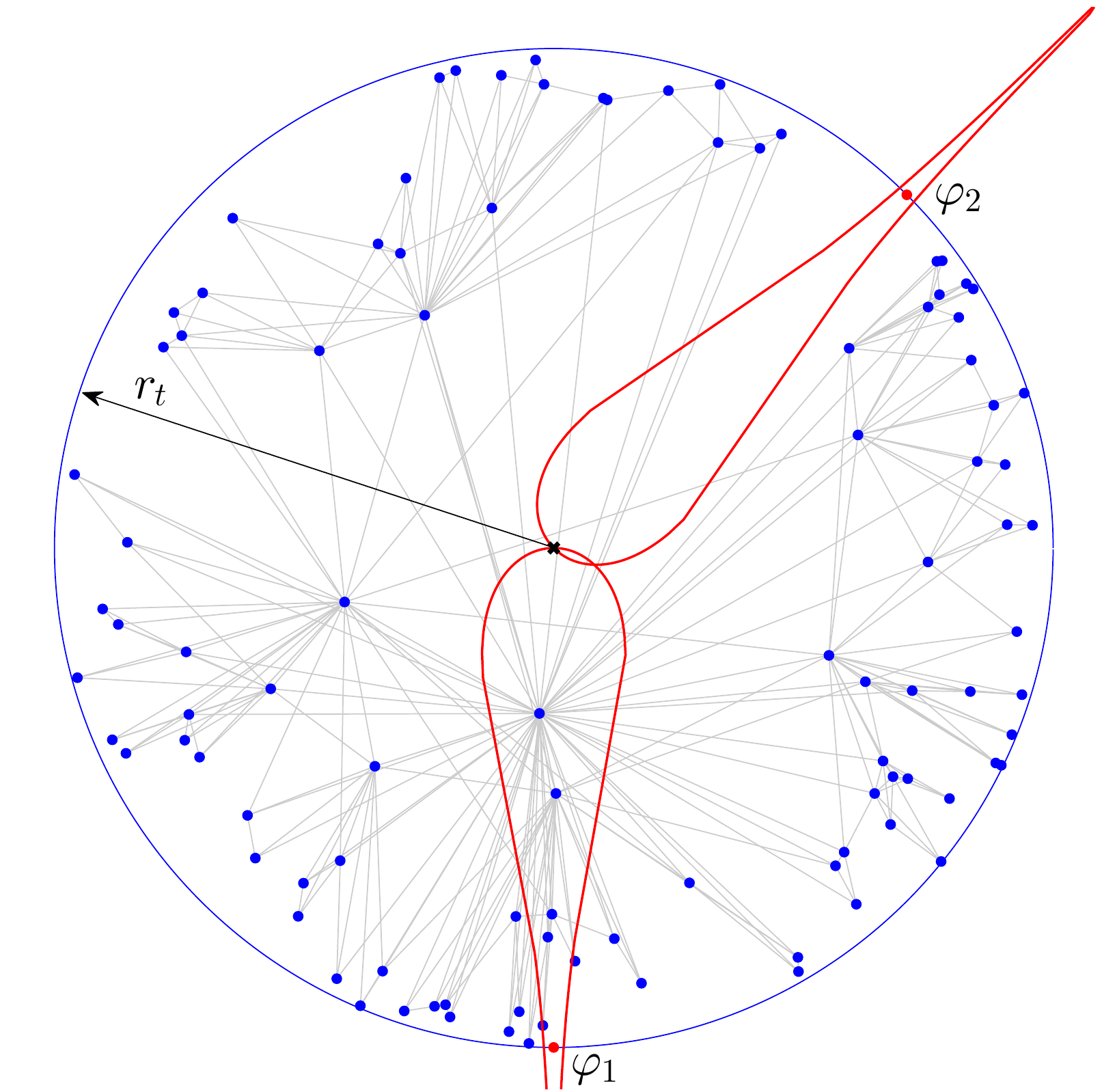}}
	\caption{\textbf{Geometric preferential attachment}. At time $t$, a new node appears at distance $r_{t}$ from the center of the hyperbolic disk denoted by cross. Points $\varphi_1$ and $\varphi_2$ represent two potential locations of the new node, and the drop-shaped curves are the boundaries of the hyperbolic disks $D_{\varphi_1}(r_t)$ and $D_{\varphi_2}(r_t)$ of radius $r_t$ centered at $\varphi_1$ and $\varphi_2$. Since similarity is attractive and $D_{\varphi_1}(r_t)$ contains more nodes (five) than $D_{\varphi_2}(r_t)$ (none), the new node is more likely to appear at $\theta_t=\varphi_1$.}
\end{figure}

The exact definition of the GPA model is:
\begin{enumerate}
	\item[0.] Initially the network is empty. New nodes $t$ appear one
	at a time, $t=1,\ldots$, and for each $t$:
	 \item[$1.$] The angular (similarity) coordinate $\theta_t$ of a new node $t$
	 is determined as follows:
    \begin{enumerate}\item[(a)] Sample $\varphi_i\sim U[0,2\pi]$, $i=1,\ldots,t$, uniformly at random. The set of points $\hat{t}_1=(r_t,\varphi_1),\ldots,\hat{t}_t=(r_t,\varphi_t)$ in the hyperbolic plane are the ``candidate'' positions for the newborn node;  
	 \item[(b)] Define the attractiveness $A_t(\varphi_i)$ of the $i^{\textrm{th}}$ candidate as the number of existing nodes that  lie within hyperbolic distance $r_t$ from it; 
	 \item[(c)] Set $\theta_t=\varphi_i$ with probability
	 \begin{equation}\label{probability}
	 \Pi_t(i)=\frac{A_t(\varphi_i)+\Lambda}{\sum_{j=1}^t(A_t(\varphi_j)+\Lambda)},
	 \end{equation}
	 where $\Lambda\geq0$ is a parameter, called the initial attractiveness.
	\end{enumerate}
    \item[2.] The radial (popularity) coordinate of node $t$ is set to $r_t=2\ln t$. The radial coordinates of existing nodes $s<t$ are updated to $r_s(t)=\beta r_s+(1-\beta)r_t$.
	\item[3.] Node $t$ connects to $m$ hyperbolically closest
	existing nodes (if $t\leq m$, then node $t$ connects to all existing nodes).
\end{enumerate}

The GPA model has thus three parameters: the number of links $m$ established by every new node, the speed of popularity fading $\beta$, and the initial attractiveness $\Lambda$. A moment's thought shows that $m$ controls the average degree of the network, $\bar{k}=2m$. We prove in Methods that the model generates scale-free networks and $\beta$ controls the power-law exponent $\gamma$. The initial attractiveness $\Lambda$ controls the heterogeneity of the angular node density, namely, the heterogeneity is a decreasing function of $\Lambda$. When $\Lambda\rightarrow\infty$, the GPA model becomes manifestly identical to the homogeneous popularity$\times$similarity (PS) model~\cite{PSO}, where the angular coordinate $\theta_t$ of a new node $t$ is sampled uniformly at random on $[0,2\pi]$. Note, however, that in GPA, choosing a position in the similarity space is an active decision made by a node based on the attractiveness of different locations, as opposed to ``passive'' uniform randomness in PS. In standard PA, the initial attractiveness term is used  to control the exponent of the power-law degree distribution~\cite{Krapivsky,Dorogovtsev2}. In what follows we show that in GPA, $\Lambda$ controls certain properties of the community size distribution.

Figure~2 shows the simulation results for networks of size $n=10^3$ generated by the GPA model with $m=3$ (i.e.\ each new node connects to the three hyperbolically closest nodes), $\beta=2/3$, and different values of $\Lambda$. As expected, the smaller the value of $\Lambda$, the more heterogeneous the distribution of angular coordinates. To quantify the difference between the empirical distribution of the angular coordinates and the uniform distribution on $[0,2\pi]$, we use the Kolmogorov-Smirnov (KS) statistic, one of the standard distances that measures the difference between two probability distributions. Recall that the KS statistic $\rho$ is defined as the maximum difference between the values of the empirical distribution $\hat{F}_n(\theta)$ of the sample $\theta_1,\ldots,\theta_n$ and the uniform distribution $F_{U[0,2\pi]}(\theta)=\theta/2\pi$,
\begin{equation}\label{KS_def}
\rho=\max_{\theta\in[0,2\pi]}\left|\hat{F}_n(\theta)-\frac{\theta}{2\pi}\right|
\end{equation}
The KS statistic as a function of $\Lambda$ is shown in the bottom panel of Fig.~2. As expected, $\rho(\Lambda)$ is a decreasing function of $\Lambda$.

\subsection{Degree Distribution}

For each of the three networks depicted in Fig.~2, the statistical procedure for quantifying power-law behavior in empirical data proposed in~\cite{powerlaws} accepts the hypothesis that the network is scale-free. It estimates the lower cutoff for the scaling region as $k_{{\rm min}}=3$, which is consistent with the minimum degree in the networks $m=3$. Figure~3(a) shows a doubly logarithmic plot of the empirical degree distributions $P(k)\sim k^{-\gamma}$ along with the fitted power-law with exponent $\gamma=2.5$.

These empirical results show that the degree distribution of a network generated by GPA appears to be a power-law. Moreover, quite unexpectedly, the power-law exponent $\gamma$ remains similar for different values of $\Lambda$. These results can be proved analytically (see Methods for details). Remarkably, for any value of $\Lambda$, the GPA model produces scale-free networks with the power-law degree distribution identical to the degree distribution in networks growing according to PA, and having power-law exponent $\gamma=1+1/\beta$.

\begin{figure}
	\centerline{\includegraphics[width=85mm]{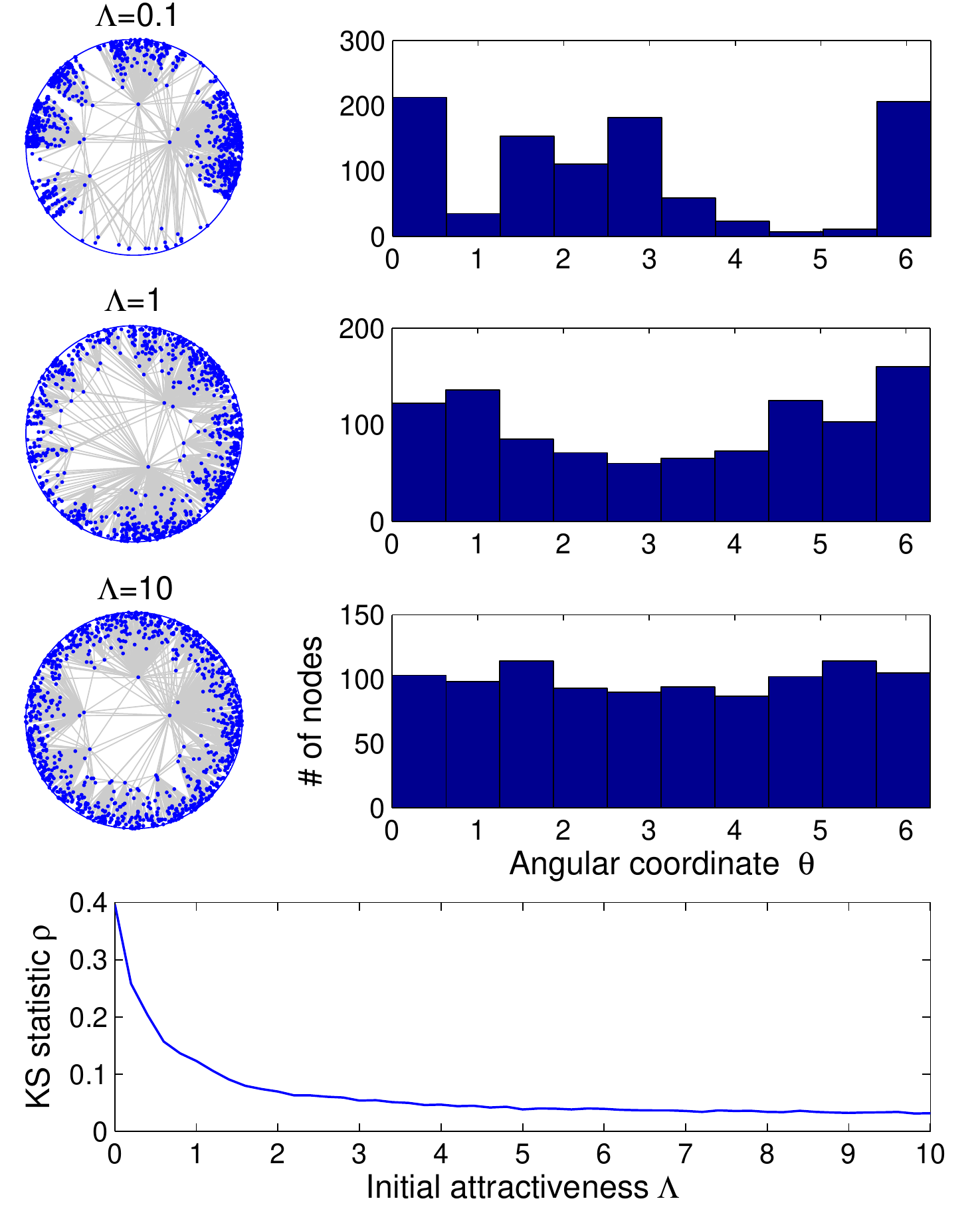}}
	\caption{\textbf{GPA networks.} Synthetic networks of size $n=10^3$ generated according to the GPA model with $m=3$, $\beta=2/3$, and $\Lambda=0.1$ (first row), $\Lambda=1$ (second row), and $\Lambda=10$ (third row). The right column shows the corresponding histograms of the angular nodes densities. The bottom panel plots the expected KS statistic $\rho$ (\ref{KS_def}), as a function of $\Lambda$. For each value of $\Lambda$, $\rho(\Lambda)$ is computed by averaging the KS statistics for $100$ independently generated networks.}
\end{figure}
\begin{figure}
	\centerline{\includegraphics[width=75mm]{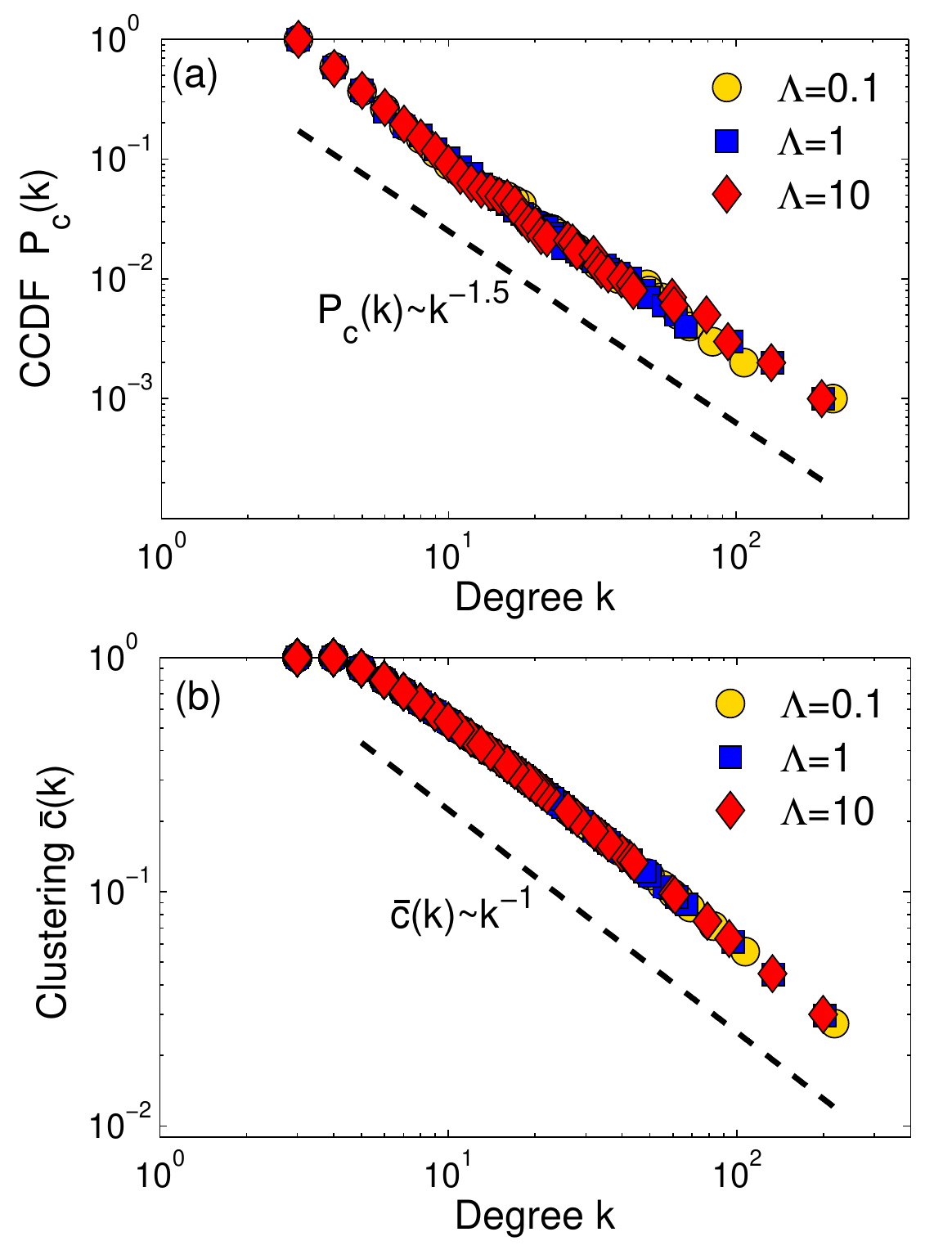}}
	\caption{\textbf{Degree distribution and clustering.} Panel (a) shows the empirical complementary cumulative degree distribution functions (CCDF) $P_c(k)=\sum_{k'=k} P(k')$ for the networks shown in Fig.~2 versus the corresponding power-law fit. The average clustering coefficient $\bar{c}(k)$ as a function of node degree $k$ for these networks is shown in panel (b). The mean clustering $\bar{c}=0.88$ for all networks.
	}
\end{figure}

\subsection{Clustering Coefficient}

The concept of clustering~\cite{smallworlds} quantifies the tendency to form
cliques (complete subgraphs) in the neighborhood of
a given node. Specifically, the local clustering coefficient of node $s$ is defined as the probability that two nodes $s'$ and $s''$, adjacent to $s$, are also connected to each other.
Figure~3(b) shows the average value of the clustering coefficient $\bar{c}(k)$ for nodes of degree $k$  as a function of $k$ for the three networks in Fig.~2. Interestingly, clustering does not depend on $\Lambda$ either (a proof is in the Methods), and scales approximately as $k^{-1}$. This means that, on average, the nodes with higher degree have lower clustering, which is consistent with empirical observations of clustering in real complex networks~\cite{Vazquez,RavaszBara}.
For all the three PGA networks, the mean clustering (the average of the local clustering coefficients) is high, $\bar{c}=0.88$.

\subsection{Soft Communities}

The hyperbolic space underlying a network and the GPA mechanism of node appearance in that space naturally induce community
structure and allow to detect communities in a very intuitive and simple way. A higher density of links within a community indicates that its nodes are more similar to each other than to the other nodes, because links connect only nodes located within a certain similarity distance threshold. All such densely linked nodes are thus close to each other in some area of the similarity space, meaning that the spatial node density is high in this area. Therefore a community becomes a cluster of spatially close nodes, and the community structure is encoded in a non-uniform distribution of angular (similarity) coordinates of nodes.

\begin{figure}
	\centerline{\includegraphics[width=85mm]{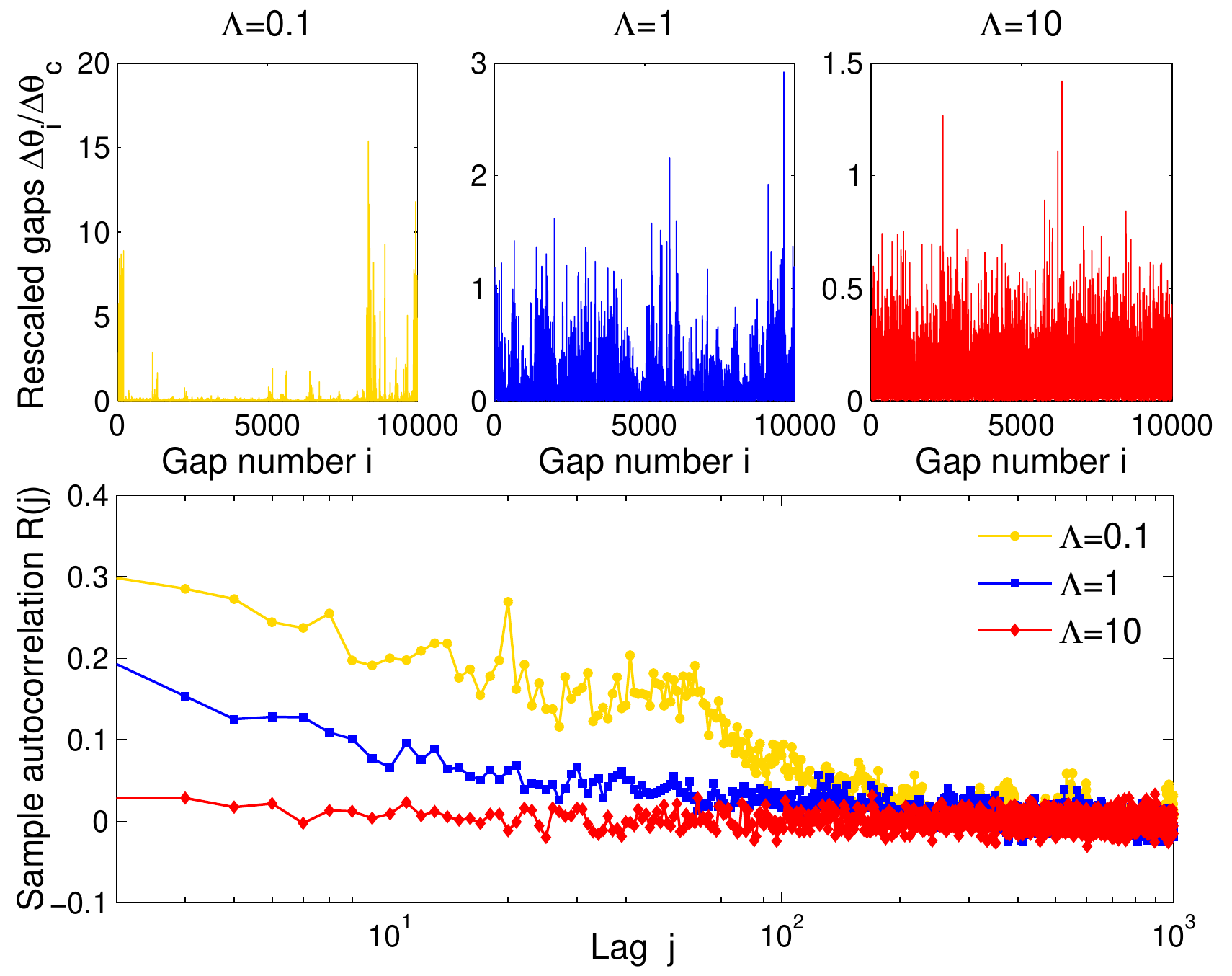}}
	\caption{\textbf{Statistics of the rescaled gaps.} Top panel shows the values of rescaled gaps $\Delta\theta/\Delta\theta_c$ for three networks of size $n=10^4$ generated by the GPA model with $\Lambda=0.1$ (left), $\Lambda=1$ (middle), and $\Lambda=10$ (right). The bottom panel shows the sample autocorrelation function of the series in the top panel.}
\end{figure}

Following the approach in~\cite{Serrano}, let us consider the angular gaps $\Delta\theta$ between consecutive nodes, and define a \textit{soft community} as a group of nodes separated from the rest of the network  by two gaps that exceed a  certain critical value $\Delta\theta_c$. If a network has a total number of $n$ nodes, then the critical gap $\Delta\theta_c$ is defined as the expected value of the largest gap $\Delta\theta_{(n)}=\max\{\Delta\theta_1,\ldots,\Delta\theta_n\}$, where $\theta_1,\ldots,\theta_n$ are distributed uniformly at random on $[0,2\pi]$. The rationale behind this definition is that if nodes are distributed uniformly in the similarity space, and there are no communities, then we do not expect to find any pair of nodes separated by a gap larger than this $\Delta\theta_c$. The calculations in the Methods show that the critical gap is approximately
\begin{equation}\label{criticalgap}
\Delta\theta_c=\frac{2\pi\ln n}{n}.
\end{equation}

Figure~4 shows the statistics of the rescaled gaps $\Delta\theta/\Delta\theta_c$ for three GPA-generated networks of size $n=10^4$ with $\Lambda=0.1, 1,$ and $10$. In the top panel, we can see the organization of nodes on the circle with many consecutive small gaps ($\Delta\theta_i<\Delta\theta_c$) indicating groups of similar nodes (communities) separated by large gaps ($\Delta\theta_i>\Delta\theta_c$) which constitute boundaries between communities, so-called ``fault lines''~\cite{Newman}. As expected, smaller values of $\Lambda$ result into more heterogeneous distribution of gaps with strong long range correlations. This effect is clearly visible in the bottom panel, where the sample autocorrelation function is shown: the smaller the $\Lambda$, the slower  the autocorrelation decays.

Having a geometric interpretation of the community structure, it is now easy to quantity how well communities are separated from each other. For each community $\mathcal{C}$, we define its \textit{separation} from the rest of the network $\mathcal{S(C)}$ as the rescaled average of two gaps $\Delta\theta_1, \Delta\theta_2>\Delta\theta_c$ that separate $\mathcal{C}$ from its neighboring communities,
\begin{equation}\label{sep}
\mathcal{S(C)}=\frac{\Delta\theta_1+\Delta\theta_2}{2\Delta\theta_c}
\end{equation}
The mean community separation, i.e.\ the expected separation of a community that a randomly chosen node belongs to, can then be computed as follows:
\begin{equation}\label{mean_sep}
\bar{\mathcal{S}}=\sum_{i=1}^{n_c} \frac{n_i}{n} \mathcal{S}(\mathcal{C}_i),
\end{equation}
where $n_i$ is the size of community $\mathcal{C}_i$ and $n_c$ is total number of communities. The network metric $\bar{\mathcal{S}}$ can also be viewed as a measure of narrowness (or specialization) of communities. For example, in scientific collaboration network, where nodes represent scientists and communities correspond to groups with similar research interests, $\bar{\mathcal{S}}$ quantifies the degree of interdisciplinarity in the network. When $\bar{\mathcal{S}}$ is large, the boundaries between communities are sharp and each community focuses on its narrow, specific topic. On the other hand, if $\bar{\mathcal{S}}$ is close to one, then the boundaries are blur, communities are wide spread, and the network is highly interdisciplinary.

The difference in the stochastic behavior of the rescaled gaps in Fig.~4 suggests that the initial attractiveness $\Lambda$ controls the mean community separation $\bar{\mathcal{S}}$ in the GPA-generated networks. This is confirmed by simulation results shown in Fig.~5(c), where $\bar{\mathcal{S}}$ is shown as a function of $\Lambda$. As expected, $\bar{\mathcal{S}}(\Lambda)$ is a monotonically decreasing function, approaching one when $\Lambda$ is large.

\subsection{The Internet}

To demonstrate the ability of the GPA mechanism to generate graphs that are similar to real networks, and, in particular, to reproduce real non-uniform distributions of similarity node coordinates, we consider the Autonomous Systems (AS) Internet topology~\cite{AS_Internet} of December 2009. The network consists of $N=25910$ nodes, ASs, and $M=63435$ links that represent  logical relationships between ASs. We embed the AS Internet into its hyperbolic space, i.e compute the popularity and similarity  coordinates $\{r_i,\theta_i\}$, using HyperMap~\cite{HyperMap}, an efficient network mapping algorithm that estimates the latent hyperbolic coordinates of nodes. The network topology has a power-law degree distribution with $\gamma=2.1$ and average node degree $\bar{k}\approx5$. This automatically determines two out of three parameters of the GPA model: $m=\bar{k}/2$ and $\beta=1/(\gamma-1)$. In Methods, we explain how to infer the value of $\Lambda$ from network data using the maximum likelihood method. 
Here we consider the snapshot of the AS Internet based on the first $n=10^3$ nodes. The corresponding estimated value of the initial attractiveness is $\Lambda_{\mathrm{Int}}=0.7$.

\begin{figure}
	\centerline{\includegraphics[width=85mm]{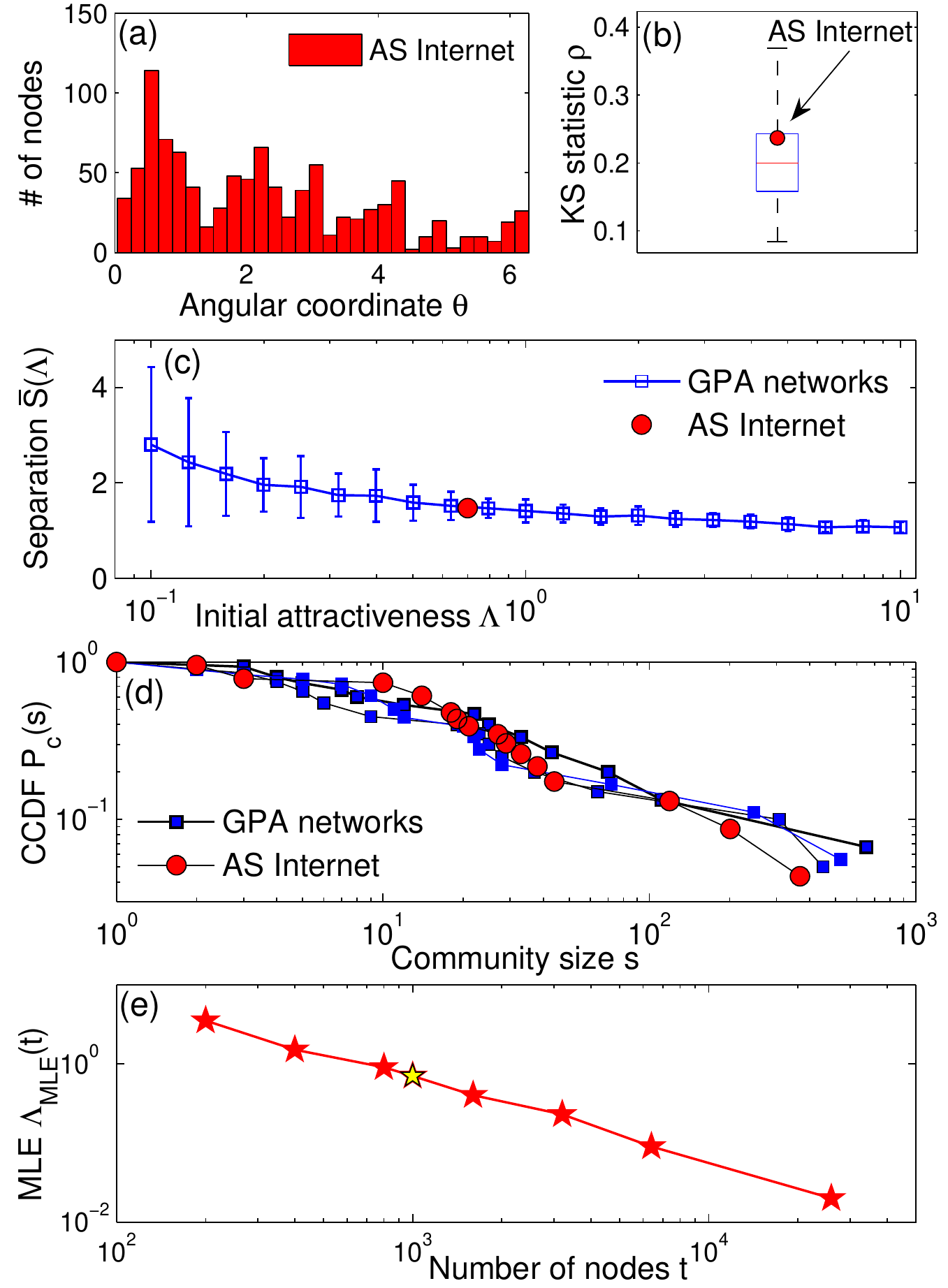}}
	\caption{\textbf{AS Internet vs GPA networks.} Panel (a) shows the histogram of the angular (similarity) coordinates $\{\theta_i\}$ for the snapshot of the AS Internet consisting of the first $n=10^3$ nodes. All $\{\theta_i\}$ are inferred by HyperMap~\cite{HyperMap}. Panel (b) compares the KS statistics for the Internet and synthetic networks generated by the GPA model (box plot) with $\gamma=2.1$ and $\Lambda=0.7$. The central red mark is the median, the blue horizontal edges of the box are the 25$^{\mathrm{th}}$ and 75$^{\mathrm{th}}$ percentiles, the black whiskers extend to the most extreme data points not considered outliers. The box plot is obtained from 100 independent generated networks. Panel (c) shows the perfect match between real and synthetic values of the mean community separation (\ref{mean_sep}). Error bars represent plus and minus one standard deviation. Panel (d) juxtaposes the empirical CCDF of the soft community sizes in the Internet against CCDFs obtained for the three GPA-generated networks. Panel (e) shows the temporal evolution of the maximum likelihood estimate $\widehat{\Lambda}_{\mathrm{MLE}}(t)$ for the AS Internet, where the node birth times are their ranks in the decreasing degree order. The yellow star corresponds to the considered snapshot with $n=10^3$ nodes and $\widehat{\Lambda}_{\mathrm{MLE}}=0.7$.}
\end{figure}

Figure~5(a) shows the histogram of the angular node density for the AS Internet snapshot. We note that it is far from uniform, which is a direct indication of the presence of soft communities. We quantify the degree of heterogeneity of the angular density by the KS distance from the uniform distribution (\ref{KS_def}) and juxtapose it against the KS distances computed for networks generated by the GPA model with $\Lambda=0.7$  (Fig.~5(b)). The Internet value lies within the 25th and 75th percentiles of the synthetic values, which shows that the degrees of non-uniformity in the Internet and GPA networks are comparable. Fig.~5(c) compares the real network with its synthetic counterpart in terms of the expected mean community separation (\ref{mean_sep}). The GPA mechanism generates networks with $\bar{\mathcal{S}}$ that match the Internet value very well. In Fig.~5(d), we compare the community size distributions in the Internet snapshot and  prediction given by the GPA model. Whereas $\bar{\mathcal{S}}$ for the Internet and GPA networks are essentially identical, the KS statistics and community size distributions are similar, but the match is not perfect. This effect is explained by the systematic bias present in the inferred values of the angular coordinates $\{\theta_i\}$. Indeed, the HyperMap method first assumes that all angular coordinates are uniformly distributed over the similarity space $\mathbb{S}^1$, i.e.\ $\Lambda=\infty$, and then perturbs them to maximize a certain likelihood function. This 	``smoothes'' the inferred angular node density and makes it more homogeneous than the true distribution. Nevertheless, although the inferred value of $\Lambda$ is only an approximation for the true value, the GPA model still captures well the degree of heterogeneity in the real network.

Finally we note that GPA defined in Eq.~\eqref{probability} admits an interesting interpretation that suggests a model extension that may be useful for real network analysis. The probability of a new node born at time $t$ to chose the angular position $\varphi_i$ can be written as
\begin{equation}
\label{probability_2}
\Pi_t(\varphi_i)=p_f \frac{A_t(\varphi_i)}{\sum_{j=1}^t A_t(\varphi_j)}+(1-p_f) \frac{1}{t},
\end{equation}
where
\begin{equation}
p_f=\frac{\langle A_t \rangle}{\langle A_t \rangle+\Lambda} \; \; \mbox{and} \; \; \langle A_t \rangle=\frac{1}{t}\sum_{j=1}^t A_t(\varphi_j).
\end{equation}
Therefore the event of choosing a position on the circle can be understood as follows. With probability $p_f$ the new node is a follower and chooses its position according to pure GPA ($\Lambda=0$). With the remaining probability $1-p_f$ the new node chooses its position uniformly at random among the $t$ available positions. We note that $\Lambda$ controls $p_f$, since $\langle A_t \rangle \approx 1$. When $\Lambda$ is constant, $p_f$ is also constant, and consequently there is always a fraction of nodes that are placed at random locations. At long times, these random nodes diminish the effect of pure GPA, and eventually the angular distribution of nodes become indistinguishable from a Poisson point process on the circle.
We can then wonder whether a constant value of $\Lambda$ is a realistic assumption for dealing with real networks.
In scientific citation networks, for example, when a new field of science is being formed, and not much work has yet been done in it, scientists may decide either to explore a new line of research within the field, or to follow one of the mainstream existing lines. The former case can be modeled by a random choice of the angular position, assuming that subfields are homogeneously distributed. The latter is modeled by the pure GPA term in Eq.~\eqref{probability_2}. However, there is a payoff that does not remain constant during the evolution of the field. At early times, the chances to find an interesting result that would be highly cited and followed by others are very high. At late times, the topic space is crowded and the chances to find something fundamentally new are very slim. Therefore, there is a higher incentive for scientists to take higher risks at early times. This can be modeled by $p_f$ increasing with time, converging to a value close to $1$ as time grows to infinity. In turn, this means that $\Lambda$ is a decreasing function of time, having a large value at the beginning of network evolution, and decreasing to small values afterwards.

Unfortunately, measuring the temporal evolution of $\Lambda$ in a real network is not yet possible because there currently exists no parametric theory describing such evolution that could be used for statistical inference of $\Lambda$. However, it is fairly easy to find an approximate value of $\Lambda$ as a function of time as follows. If timestamps of a real complex network are available, we can pretend that $\Lambda$ is constant, and infer its value using the MLE techniques described in the Methods for subgraphs made of nodes that were born before a given time $t$, $\widehat{\Lambda}_{\mathrm{MLE}}(t)$. This value can be thought of as a (possibly weighted) average of $\Lambda(t)$ in time window $(0,t)$. By increasing the value of $t$, we can detect whether $\Lambda$ is constant (if $\widehat{\Lambda}_{\mathrm{MLE}}(t)$ does not change with time, beyond statistical fluctuations), or a decreasing function of time. Figure~5(e) shows $\widehat{\Lambda}_{\mathrm{MLE}}(t)$ for the AS Internet where the strong temporal dependence of $\Lambda$ is evident.

\section{Discussion}
In summary, hyperbolic network geometry, combining popularity and similarity forces driving network evolution, and coupled with preferential attachment of nodes to this geometry (GPA), naturally yields scale-free, strongly clustered growing networks with emergent soft community structure. The GPA model has three parameters that can be readily inferred from network data. Using the AS Internet topology as example, we have seen that the GPA mechanism generates heterogeneous networks that are similar to real networks with respect to key properties, including key aspects of the community size distribution and separation. The mean community separation, a new metric that quantifies the narrowness of communities in a network, is controlled in GPA by initial attractiveness $\Lambda$, which controls the power-law exponent in standard PA.

In the context of the asymptotic equivalence between de Sitter causal sets and popularity$\times$similarity (PS) hyperbolic networks established in~\cite{NetCosm}, we note that $\Lambda$ is conceptually similar to the cosmological constant $\Lambda$ in Einstein's equations in general relativity (GR), where it is also an additive term in the proportionality between the energy-momentum tensor and spacetime curvature. Causal sets~\cite{Bombelli,Dowker} are random graphs obtained by Poisson sprinkling a collection of nodes onto (a patch) of a Lorentzian manifold; edges in these graphs connect all timelike-separated pairs of nodes. If there is no matter (empty spacetime) but there is only dark energy (positive $\Lambda$), then the solution of Einstein's equations is the de Sitter spacetime, and the main theorem in~\cite{NetCosm} states that the ensemble of PS graphs is asymptotically ($n\to\infty$) identical to an ensemble of causal sets sprinkled onto de Sitter spacetime, which is one of the three maximally symmetric, homogeneous and isotropic Lorentzian manifolds (the other two are Minkowski and anti-de Sitter spacetimes). In this context, the GPA model considered here is a model with cosmological constant $\Lambda$ and matter. Modeled by high node density, this matter, as in GR, ``attracts more matter,'' thus increasing the spacetime curvature of which the node density is a proxy. Indeed the main feature of the model is that the higher the node density in a particular region of space, the more nodes will appear in this region later. The main difference with GR is that here we essentially have an analogy with only the $00$-component of Einstein's equations. One can envision that other components should describe the coupled dynamics of the similarity space and nodes in it. In case of scientific collaboration network, for example, that would be the co-evolution of science (space) itself, and interests of scientists (node dynamics in this space). In the model considered here nodes do not move. Finding the laws of their spatial dynamics that may further strengthen the analogy with general relativity is a promising but challenging research direction.

In that context, the decay of initial attractiveness $\Lambda$ that we found in the Internet must be analogous to the decay of cosmological constant $\Lambda$ in modern cosmological theories. Cosmic inflation~\cite{Lyth,Mukhanov} is widely accepted as the most plausible resolution of many problems with the classical big bang theory, including the flatness problem, the horizon problem, and the magnetic-monopole problem. Inflation is an initial period of accelerated expansion of the universe during which gravity was repulsive. Inflation does not last long, and can be modeled as a time dependent cosmological ``constant'' $\Lambda$ that initially has a high value and then decays to zero. The analogies between GPA with decaying $\Lambda$ and inflation go even further, producing similar outcomes as far as the spatial distribution of events is concerned. Indeed, cosmic inflation has the effect of smoothing out inhomogeneities so that once inflation is over, the universe is nearly flat, isotropic, and homogeneous, except for quantum fluctuations of the inflaton field. These fluctuations are the seeds of future inhomogeneities that we observe in the universe at scales smaller than 100 Mpc. In the GPA context, a high value of $\Lambda$ has also a homogenizing effect. Indeed, if $\Lambda$ is large, then $p_f$ is small, and new nodes chose their angular positions at random, producing a Poisson point process on the circle. Once $\Lambda$ is small enough, we are left with a random distribution of points with Poisson fluctuations that, as in the universe, are the seeds of future communities in the network (galaxies in the universe), because once $\Lambda$ is nearly zero, these initial fluctuations are reinforced by pure preferential attachment.

\section{Methods}

\subsection{Invariance of the degree distribution and clustering}

Here we prove that the degree distribution and clustering coefficient in the networks generated by the GPA model do not depend of the initial attractiveness $\Lambda$. Moreover, the degree distribution is power-law with exponent $\gamma=1+1/\beta$. The proof can be reduced to the proof for the homogeneous PS model~\cite{PSO} (Supplementary Information, Section IV). Consider a new node $t$, and let
$R_t$ be the radius of a hyperbolic disk centered at this node such
that $t$ is connected to all nodes $s<t$ that lie in this disc. Then the probability $\mathbb{P}_{\rm{GPA}}(s,t)$ that nodes $t$ and $s<t$ in the GPA model are connected can be computed as follows:
\begin{equation}\label{link prob1}
\begin{split}
\mathbb{P}_{\rm{GPA}}(s,t)&=\mathbb{P}(x_{st}\leq R_t)\\
&= \mathbb{P}\left(\theta_{st}\leq 2e^{-\frac{r_s(t)+r_t-R_t}{2}}\right),
\end{split}
\end{equation}
where $x_{st}=r_s(t)+r_t+2\ln\frac{\theta_{st}}{2}$ is the hyperbolic distance between nodes $s=(r_s(t),\theta_s)$ and $t=(r_t,\theta_t)$ at time $t$. Using the total probability theorem,
\begin{equation}\label{link prob2}
\begin{split}
\mathbb{P}_{\rm{GPA}}(s,t)=&
\sum_{i=1}^t\mathbb{P}\left(\left.\theta_{st}\leq
2e^{-\frac{r_s(t)+r_t-R_t}{2}}\right| t=\hat{t}_i\right)\mathbb{P}(t=\hat{t}_i) \\
=&\sum_{i=1}^t\mathbb{P}\left(\theta_{s\hat{t}_i}\leq
2e^{-\frac{r_s(t)+r_t-R_t}{2}}\right)\Pi_t(i),
\end{split}
\end{equation}
where $\hat{t}_i$ are the candidate positions generated at Step $1(a)$, and $\Pi_t(i)$ are the corresponding acceptance probabilities (\ref{probability}). Applying the total probability theorem with respect to node $s$, we have:
\begin{equation}\label{link prob3}
\begin{split}
&\mathbb{P}_{\rm{GPA}}(s,t)=\\
&=\sum_{i=1}^t\sum_{j=1}^s\mathbb{P}\left(\left.\theta_{s\hat{t}_i}\leq
2e^{-\frac{r_s(t)+r_t-R_t}{2}}\right|s=\hat{s}_j\right)\mathbb{P}(s=\hat{s}_j)\Pi_t(i)\\
&=\sum_{i=1}^t\sum_{j=1}^s\mathbb{P}\left(\theta_{\hat{s}_j\hat{t}_i}\leq
2e^{-\frac{r_s(t)+r_t-R_t}{2}}\right)\Pi_s(j)\Pi_t(i)
\end{split}
\end{equation}
Since the angular coordinates of the candidate positions $\hat{s}_j$ and $\hat{t}_i$ are uniformly distributed on $[0,2\pi]$, the probability $\mathbb{P}(\theta_{\hat{s}_j\hat{t}_i}\leq\alpha)$ is simply $\alpha/\pi$. Therefore,
\begin{equation}\label{link prob4}
\begin{split}
\mathbb{P}_{\rm{GPA}}(s,t)=&\frac{2}{\pi}e^{-\frac{r_s(t)+r_t-R_t}{2}}\sum_{i=1}^t\Pi_t(i)\sum_{j=1}^s\Pi_s(j)\\
=&\frac{2}{\pi}e^{-\frac{r_s(t)+r_t-R_t}{2}},
\end{split}
\end{equation}
where the last equality holds because $\sum_{i=1}^t\Pi_t(i)=\sum_{j=1}^s\Pi_s(j)=1$. We note that $\mathbb{P}_{\rm{GPA}}(s,t)$ does not depend on $\Lambda$, and that it is exactly the same as the probability $\mathbb{P}_{\rm{PS}}(s,t)$ of having a link between nodes $t$ and $s<t$ in the homogeneous PS model. The rest of the proof repeats
the proof in~\cite{PSO} without a change. This  leads to
\begin{equation}\label{PSO_L=PSO=PA}
\mathbb{P}_{\rm{GPA}}(s,t)=\mathbb{P}_{\rm{PS}}(s,t)=\mathbb{P}_{\rm{PA}}(s,t)=m\frac{\left(\frac{s}{t}\right)^{-\beta}}{\int_1^t\left(\frac{s}{t}\right)^{-\beta}ds},
\end{equation}
which means that the resulting degree distribution in
GPA is identical to PA: it is the power-law with exponent $\gamma=1+1/\beta$. Since the connection probability  $\mathbb{P}_{\rm{GPA}}(s,t)$  does not depend on $\Lambda$, neither does clustering.

\subsection{Critical gap}

To obtain a closed-form expression for the critical gap, we note that for large $n$, the sequence $\theta_1,\ldots,\theta_n\sim U[0,2\pi]$ can be  approximately viewed as a realization of the Poisson point process on the circle of unit radius with density $\lambda=n/2\pi$. In this case, the distribution of the angular gaps is approximately exponential with rate $\lambda$. The maximum gap $\Delta\theta_{(n)}$ has then the following PDF
$f_{\Delta\theta_{(n)}}(x)=
\frac{n^2}{2\pi}e^{-\frac{n}{2\pi}x}\left(1-e^{-\frac{n}{2\pi}x}\right)^{n-1}$,
and its expected value can be calculated as follows:
\begin{equation}
\begin{split}
\Delta\theta_c=&~\frac{n^2}{2\pi}\int_0^{\infty}xe^{-\frac{n}{2\pi}x}\left(1-e^{-\frac{n}{2\pi}x}\right)^{n-1}dx\\
=&-2\pi\int_0^1y^{n-1}\ln(1-y)dy\\
=&2\pi\int_0^1y^{n-1}\sum_{k=1}^{\infty}\frac{y^k}{k}dy=2\pi\sum_{k=1}^{\infty}\frac{1}{k(n+k)}\\
=&\frac{2\pi H_n}{n}\approx\frac{2\pi(\ln n +\gamma)}{n}\approx \frac{2\pi\ln n}{n},
\end{split}
\end{equation}
where $H_n$ is the $n^{\mathrm{th}}$ harmonic number, and $\gamma$ is Euler's constant.

\subsection{Inference of $\Lambda$}

The initial attractiveness $\Lambda$ controls the distribution of angular coordinates $\theta_1,\ldots,\theta_n$ of the nodes. We therefore first infer $\theta_i$ using the HyperMap method~\cite{HyperMap}.  Given the network embedding $\{(r_i,\theta_i)\}_{i=1}^n$ into its hyperbolic space, the likelihood function $\mathcal{L}(\Lambda|\theta_1,\ldots,\theta_n)$ can be written as follows:
\begin{equation}\label{Liklihood}
\begin{split}
&\mathcal{L}(\Lambda|\theta_1,\ldots,\theta_n)=\mathbb{P}(\theta_1,\ldots,\theta_n|\Lambda)\\
&=\mathbb{P}(\theta_1|\Lambda)\mathbb{P}(\theta_2|\Lambda,\theta_1)\ldots\mathbb{P}(\theta_n|\Lambda,\theta_1,\ldots,\theta_{n-1})\\
&\propto\int\limits_0^{2\pi}\frac{(A_2(\theta_2)+\Lambda)d\varphi_1}{A_2(\theta_2)+A_2(\varphi_1)+2\Lambda}\times\ldots \\
&\times\int\limits_0^{2\pi}\ldots\int\limits_0^{2\pi}\frac{(A_n(\theta_n)+\Lambda)d\varphi_1\ldots d\varphi_{n-1}}{A_n(\theta_n)+\sum_{i=1}^{n-1}A_n(\varphi_i)+n\Lambda},\\
\end{split}
\end{equation}
where $A_t(\varphi)$ is the attractiveness of location $\varphi\in\mathbb{S}^1$, 
that is the number of existing nodes at time $(t-1)$ that lie within distance $r_t$ from $(r_t,\varphi)$. The log-likelihood is then (up to an additive constant):
\begin{equation}\label{logLiklihood}
\begin{split}
&l(\Lambda|\theta_1,\ldots,\theta_n)=\\
&=\sum\limits_{t=2}^n\log\int\limits_0^{2 \pi}\ldots\int\limits_0^{2\pi}\frac{(A_t(\theta_t)+\Lambda)d\varphi_1\ldots d\varphi_{t-1}}{A_t(\theta_t)+\sum_{i=1}^{t-1}A_t(\varphi_i)+t\Lambda}
\end{split}
\end{equation}
The multiple integrals in (\ref{logLiklihood}) cannot be calculated analytically, since the attractiveness function cannot be written in closed-form. Nevertheless, the log-likelihood can be efficiently estimated be the Monte Carlo method. First, generate $N$ Monte Carlo samples, $\varphi_1^{(j)},\ldots,\varphi_{n-1}^{(j)}\sim U[0,2\pi]$, $j=1,\ldots,N$. The ``truncated'' samples  $\varphi_1^{(j)},\ldots,\varphi_{t-1}^{(j)}$  will be used for estimating the $(t-1)$-dimensional integral in (\ref{logLiklihood}). Next, precompute all needed attractivenesses, $A_t(\varphi_i^{(j)})$, where $t=2,\ldots,n$ and $i=1,\ldots,t-1$.  Then for each value of $\Lambda$, the log-likelihood can be
estimated as follows (up to a constant):
\begin{equation}\label{LogLikMC}
\begin{split}
&l(\Lambda|\theta_1,\ldots,\theta_n)\approx\\
&\approx\sum\limits_{t=2}^n\log\left(\frac{1}{N}\sum\limits_{j=1}^N\frac{A_t(\theta_t)+\Lambda}{A_t(\theta_t)+\sum_{i=1}^{t-1}A_t(\varphi_i^{(j)})+t\Lambda}\right)
\end{split}
\end{equation}

\begin{table}
	\centerline{
	\begin{tabular}{l c c c c c c}
		\hline\hline
		True $\Lambda$	& $0$ & $0.2$ & $0.5$ & $0.7$ & $1$ & $2$ \\ [0.5ex] 
		\hline
		$\widehat{\Lambda}_{\mathrm{MLE}}(100)$& 0& 0.3 & 0.4 & 0.7 & 0.7 & 1.4  \\
		$\widehat{\Lambda}_{\mathrm{MLE}}(200)$& 0& 0.2 & 0.5 & 0.8 & 1.3 & 1.8  \\
		$\widehat{\Lambda}_{\mathrm{MLE}}(500)$& 0& 0.2 & 0.6 & 0.7 & 1.1 & 1.9 \\
		$\widehat{\Lambda}_{\mathrm{MLE}}(1000)$& 0& 0.2 & 0.5 & 0.7 & 1.1 & 1.8  \\
		\hline
	\end{tabular}}
		\caption{\textbf{Maximum likelihood estimates.} True values of the initial attractiveness parameter $\Lambda$ and its MLEs $\widehat{\Lambda}_{\mathrm{MLE}}(n_0)$ based on the first $n_0=100, 200, 500,$ and $1000$ nodes. In all simulations, $N=100$ Monte Carlo samples were used in (\ref{LogLikMC}).}
\end{table}
\begin{figure}
	\centerline{\includegraphics[width=80mm]{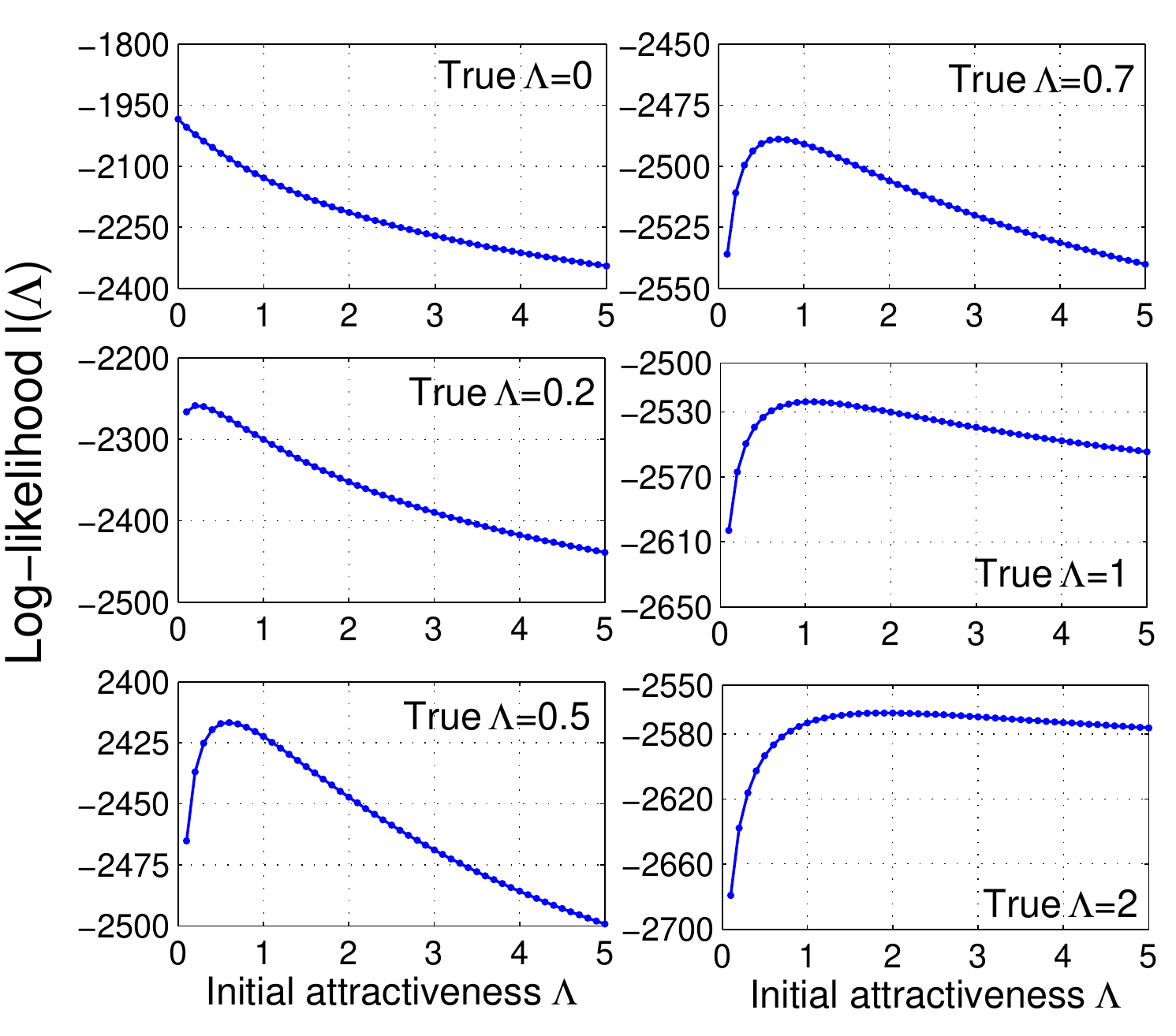}}
	\caption{\textbf{Log-likelihood functions.} The estimated log-likelihood functions $l(\Lambda|\theta_1,\ldots,\theta_n)$ for synthetic networks of size $n=10^3$ generated by the GPA model with $\Lambda=0, 0.2, 0.5, 0.7, 1,$ and $2$. Each log-likelihood is estimated by (\ref{LogLikMC}) using $N=100$ Monte Carlo samples and $n_0=500$ first nodes.}
\end{figure}

Computing attractivenesses of the Monte Carlo samples $A_t(\varphi_i^{(j)})$ involves computing $O(n^3N)$ hyperbolic distances, which is the most computationally intensive part of the algorithm. Having all attractivenesses computed, we can then estimate $l(\Lambda)$ for any $\Lambda\in[0:\Delta\Lambda:\Lambda_{\mathrm{max}}]$, and find the maximum likelihood estimate (MLE) $\widehat{\Lambda}_{\mathrm{MLE}}$. An important observation that drastically improves the efficiency of the algorithm is that we do not have to use the entire network to accurately estimate  $\widehat{\Lambda}_{\mathrm{MLE}}$, the first $n_0\ll n$ nodes are often enough. Table~1 shows the MLEs $\widehat{\Lambda}_{\mathrm{MLE}}(n_0)$ obtained from the first $n_0=100, 200, 500$, and $1000$ nodes of the networks generated by the GPA model with  $\Lambda=0, 0.2, 0.5, 0.7, 1,$ and $2$. The corresponding log-likelihood functions are shown in Fig.~6. These simulation results show that  the smaller the true value of $\Lambda$ ---  and we expect it to be small in real networks since most of them have community structure --- the less network data we need to pin $\widehat{\Lambda}_{\mathrm{MLE}}$ down. If, for example, $\Lambda=0$, then the MLE of $\Lambda$  based on the first $n_0=100$ nodes is already zero. The larger the true value of $\Lambda$, however, the flatter the log-likelihood is around its maximum, which makes inference more challenging.

\begin{acknowledgments}
This work was supported by DARPA grant No.\ HR0011-12-1-0012; NSF grants No.\ CNS-1344289,
CNS-1442999, CNS-0964236, CNS-1441828, CNS-1039646, and CNS-1345286; by Cisco Systems; by the James S. McDonnell Foundation Scholar Award in Complex Systems; by the Icrea Foundation, funded by the {\it Generalitat de Catalunya}; by the MINECO project No.\ FIS2013-47282-C2-1-P; and by the {\it Generalitat de Catalunya} grant No.\ 2014SGR608.
\end{acknowledgments}

\end{document}